\documentclass[letter,longauth]{aa}
\usepackage{graphicx}
\usepackage{tikz}
\usepackage{gensymb}
\usepackage{titlesec} 
\usepackage{placeins} 
\usepackage{txfonts}
\newcommand{\as}{$^{\prime\prime}$}

\usepackage{float}

\usepackage[hidelinks,colorlinks=true,linkcolor=blue,citecolor=blue]{hyperref}

\begin{document} 

\title{J-PAS: Discovery of RaJav, a bright extended Ly$\alpha$ Nebula at z=2.25}

\titlerunning{The discovery of the RaJav nebula}
\authorrunning{Rahna et al.}
\author{P. T. Rahna\inst{\ref{CEFCA} \thanks{rahna.payyasseri@gmail.com}}
\and M. Akhlaghi\inst{\ref{CEFCA},\ref{CEFCA2}}
\and J. A. Fernández-Ontiveros\inst{\ref{CEFCA},\ref{CEFCA2}}       
\and Z. -Y. Zheng\inst{\ref{SHAO}}
\and A. Hernán-Caballero\inst{\ref{CEFCA},\ref{CEFCA2}}
\and R. Amorín\inst{\ref{IAA}}
\and C. L\'opez-Sanjuan\inst{\ref{CEFCA},\ref{CEFCA2}}
\and J. M. Diego \inst{\ref{IFCA}}
\and L.A. D\'iaz-Garc\'ia \inst{\ref{IAA}}
\and J. M. Vílchez \inst{\ref{IAA}}
\and A. Lumbreras-Calle \inst{\ref{CEFCA}}
\and D. Fernández Gil\inst{\ref{CEFCA}}
\and S. Gurung-López \inst{\ref{OVA}}
\and Y. Jiménez-Teja \inst{\ref{IAA},\ref{ON}}
\and A.~Ederoclite\inst{\ref{CEFCA},\ref{CEFCA2}}
\and R. M. González Delgado\inst{\ref{IAA}}
\and H.~V\'azquez Rami\'o\inst{\ref{CEFCA},\ref{CEFCA2}}
\and R.~Abramo\inst{\ref{IF/USP}}
\and J.~Alcaniz\inst{\ref{ON}}
\and N.~Ben\'itez\inst{\ref{IAA}}
\and S.~Bonoli\inst{\ref{DIPC},\ref{Iker}}
\and S.~Carneiro\inst{\ref{ON}}
\and J.~Cenarro\inst{\ref{CEFCA},\ref{CEFCA2}}
\and D.~Crist\'obal-Hornillos\inst{\ref{CEFCA}}
\and R.~Dupke\inst{\ref{ON}}
\and C.~Hern\'andez-Monteagudo\inst{\ref{ULL},\ref{IAC}}
\and A.~Mar\'in-Franch\inst{\ref{CEFCA},\ref{CEFCA2}}
\and C.~Mendes de Oliveira\inst{\ref{SaoPaulo}}
\and M.~Moles\inst{\ref{CEFCA},\ref{IAA}}
\and L.~Sodr\'e Jr.\inst{\ref{SaoPaulo}}
\and K.~Taylor\inst{\ref{Inst4}}
\and J.~Varela\inst{\ref{CEFCA}}
}          
\institute{Centro de Estudios de F\'isica del Cosmos de Aragón (CEFCA), Plaza San Juan 1, 44001 Teruel, Spain\label{CEFCA}
\and Unidad Asociada CEFCA-IAA, CEFCA, Unidad Asociada al CSIC por el IAA y el IFCA, Plaza San Juan 1, 44001 Teruel, Spain\label{CEFCA2}
\and Key Laboratory for Research in Galaxies and Cosmology, Shanghai Astronomical Observatory, CAS, Shanghai 200030, PR China\label{SHAO}
\and Instituto de Astrof\'{\i}sica de Andaluc\'{\i}a (CSIC), P.O.~Box 3004, E-18080 Granada, Spain\label{IAA}
\and Instituto de Física de Cantabria, Edificio Juan Jordá, Avenida de los
Castros, 39005 Santander, Spain \label{IFCA}
\and Observatori Astron\`omic de la Universitat de Val\`encia, 46980 Paterna, Spain  \label{OVA}
\and Instituto de F\'isica, Universidade de S\~ao Paulo, Rua do Mat\~ao 1371, CEP 05508-090, S\~ao Paulo, Brazil\label{IF/USP}
\and Observat\'orio Nacional -- MCTI (ON), Rua General Jos\'e Cristino, 77, S\~ao Crist\'ov\~ao, 20921-400, Rio de Janeiro, Brazil\label{ON}
\and Donostia International Physics Centre, Paseo Manuel de Lardizabal 4, E-20018 Donostia-San Sebasti\'an, Spain\label{DIPC}
\and Ikerbasque, Basque Foundation for Science, E-48013 Bilbao, Spain\label{Iker}
\and Departamento de Astrof\'isica, Universidad de La Laguna, E-38206, La Laguna, Spain\label{ULL}
\and Instituto de Astrof\'isica de Canarias, E-38200 La Laguna, Spain\label{IAC} 
\and Departamento de Astronomia, Instituto de Astronomia, Geof\'isica e Ci\^encias Atmosf\'ericas da USP, Cidade Universit\'aria, 05508-900, S\~ao Paulo, SP, Brazil\label{SaoPaulo}
\and Instruments4, 4121 Pembury Place, La Ca\~nada-Flintridge, CA, 91011, USA\label{Inst4}
}
  \abstract
   {We report the discovery of a massive and potentially largest Ly$\alpha$ Nebula, RaJav, at z=2.25, associated with a quasar pair: the bright SDSS~J162029.07+433451.1 (hereafter J1620+4334) and the faint newly discovered quasar JPAS-9600-10844, at 2.265 $\pm$ 0.021 using the early data release (17 deg$^{2}$) of the Javalambre Physics of the Accelerating Universe Astrophysical Survey (J-PAS). 
   The quasar JPAS-9600-10844 embedded in the nebula is located at $\sim$ 60.2 kpc (7.3\as) from J1620+4334, and shows a compact structure with broad emission lines (> 3000 km/s), typical of active galactic nuclei (i.e., Ly$\alpha$ $\lambda$1216 and CIV $\lambda$1548). At a 2$\sigma$ surface brightness (SB) contour of $\sim 1.86 \times 10^{-16}$ erg s$^{-1}$ cm$^{-2}$ arcsec$^{-2}$, the nebula extends > 100 kpcs and has a total Ly$\alpha$ luminosity of $\sim 5.8 \pm 0.7 \times 10^{44}$ erg s$^{-1}$ signify the presence of a giant Enormous Ly$\alpha$ Nebula (ELAN). The nebula traces an over density of quasars at redshift of 2.2-2.3 consistent with the progenitor of a massive galaxy cluster. The extended CIV emission with luminosity of $\sim 3.7 \pm 0.5 \times 10^{44}$ erg s$^{-1}$ indicates that the circum-galactic medium (CGM) is metal-enriched and not primordial.
   The current J-PAS observations suggest photoionization and shocks due to outflows as possible ionization mechanisms. The faint extended FUV and NUV continuum emission likely points to ongoing star formation around the two quasars, suggesting a complex interaction in their environments.
   These findings provide new insights into the environment of quasars and their role in shaping the dynamics and evolution of the CGM at cosmic noon. Further spectroscopic observations will be required to fully characterize the object's nature and its kinematic properties. This study demonstrates the unique capability of J-PAS to detect massive and rare Ly$\alpha$ nebulae, providing new insights into their properties, environments, and connections to large-scale structures in the cosmic web such as filaments and overdensities in a large cosmological volume.}
  
   \keywords{cosmology: observations --
                galaxy: formation --
                quasars: general
               }

\maketitle
%
\begin{figure*}
\centering
\includegraphics[angle=0,width=18.5cm]{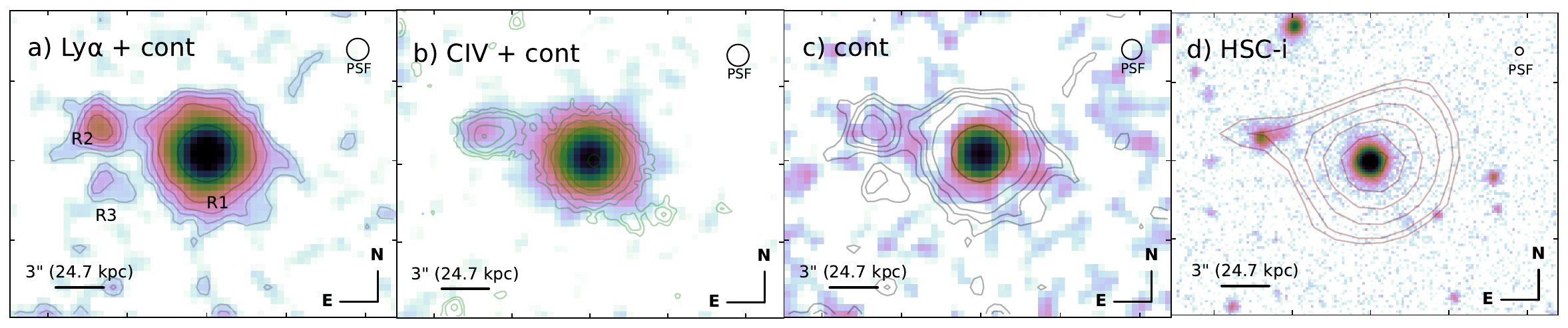}
\caption{Co-added J-PAS NB image of a) Ly$\alpha$ +continnum (J0390+J0400+J0410), b) CIV +continnum (J0490+J0500+J0510), and c) continuum (J0420+J0440+J0450). Broad band image from d) HSC i-band. SB contours in the panels a and c are from the Ly$\alpha$ image (levels: 1.3$\sigma$, 2$\sigma$,3$\sigma$,10$\sigma$, and 100$\sigma$), in panel b is from HSC g-band (partially including Ly$\alpha$ and CIV) image (levels: 2$\sigma$,3$\sigma$,5$\sigma$,10$\sigma$,20$\sigma$ and 1000$\sigma$) and the panel d overlaid with radio LOFAR 144 MHz contours (levels: 2$\sigma$,2.5$\sigma$,4$\sigma$, 6$\sigma$, 8$\sigma$, and 9.5$\sigma$). The positions of the bright quasar J1620+4334, the faint quasar JPAS-9600-10844, and the Ly$\alpha$ emitting region are labeled as R1, R2, and R3, respectively and RaJav nebula defined in blue contour.}
\label{Fig:jpasim}
\end{figure*}
\begin{figure}
\centering
\includegraphics[angle=0,width=9cm]{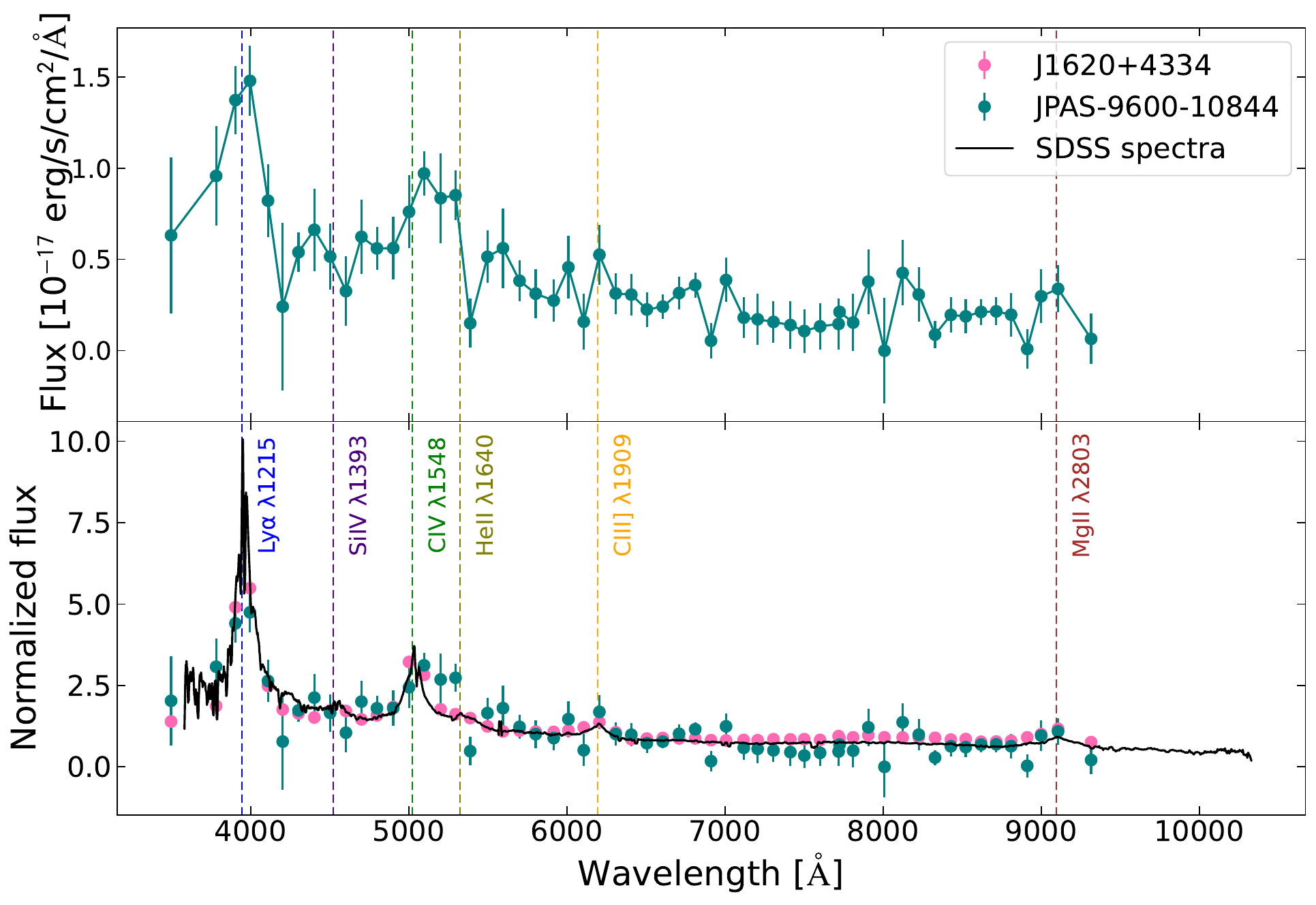}
\caption{Upper panel: J-spectra of JPAS-9600-10844. Bottom panel: Normalized photospectra of JPAS-9600-10844 and J1620+4334, along with the normalized SDSS spectrum of J1620+4334.}
\label{Fig:jspec}
\end{figure}
\section{Introduction}
The circum-galactic medium (CGM) is a region of diffuse gas that surrounds the galaxies within the virial radii of their dark-matter halos, and plays a crucial role in the formation and evolution of the structures in the early Universe. The Ly$\alpha$ $\lambda$1216 emission line serves as a key tracer of the ionized CGM at high redshifts. The ultraviolet (UV) radiation from galaxies and quasars \citep{2006Adelberger, 2012Cantalupo} ionize the surrounding gas and produces Ly$\alpha$ nebulae. These nebulae allow us to understand gas accretion, outflows, and the relation between galaxies and the intergalactic medium.

Several studies have analyzed luminous spatially extended Ly$\alpha$ nebulae or halos around high-z (z = 2 to 7) quasars \citep[e.g.,][]{1987Hu,2014Cantalupo, 2015Hennawi, 2019Farina, 2022Rahna, 2024Li} in the last two decades. These diffuse Ly$\alpha$ nebulae are luminous (> 10$^{43}$ erg s$^{-1}$ ), extended (> 40kpc) and very faint (surface brightness (SB) of  $\sim 10^{-18}$ to $~10^{-20} $ erg s$^{-1}$ cm$^{-2}$ arcsec$^{-2}$). The origin of these nebulae is still under debate and the possible powering mechanisms are photo-ionization from quasars, super wind-driven shocks, gravitational cooling radiation from cold accretion, and resonant scattering of quasar radiation  \citep[e.g.,][]{2000Haiman,2016Trebitsch,2017Cantalupo,2000Taniguchi}. Deep observations (SB$_{Ly\alpha} < 10^{-18} $ erg s$^{-1}$ cm$^{-2}$ arcsec$^{-2}$) using large telescopes (e.g., MUSE/VLA, Keck) were able to detect extreme cases of enormous Ly$\alpha$ nebulae (ELANe) such as UM287 nebula, Jackpot, MAMMOTH-1, and Fabulous (with sizes from 300 to 500 kpcs), in the vicinity of high luminous quasars at cosmic noon (z=2-3) over the last decade \citep{2014Cantalupo,2015Hennawi,2017Cai, 2018Arrigoni}. 
ELANe are luminous nebulae that trace galaxy density peaks at high redshift, often found at filament intersections marking the nodes of the cosmic web. Their size, rarity, and association with dense environments make them powerful tools for studying galaxy formation, gas accretion, and to probe the large-scale structure in the early universe. However, most of the Ly$\alpha$ nebulae discoveries to date rely on pre-selection methods applied to small fields of view, constrained to narrow redshift intervals and using a limited set of narrow-band (NB) filters. This may result in an incomplete picture of the overall Ly$\alpha$ nebulae population and its properties. Providing a systematic census of these massive structures is essential to understand their role in galaxy formation and evolution, as well as their impact on the large-scale cosmic structure.
 
In contrast, the Javalambre Physics of the Accelerating Universe Astrophysical Survey \citep[J-PAS;][] {2014Benitez}, a wide-field, multi-narrowband survey offers a complementary opportunity to identify the Ly$\alpha$ nebulae population \citep[e.g., ][]{2022Rahna} in a large area (8500 deg$^{2}$ by its completion). It uses a 3deg$^{2}$ camera to observe the sky with 54 NBs (140 \AA~width) filters, 2 medium bands (on the extreme red/blue) and one broad band (iSDSS). J-PAS delivers low-resolution spectra (spectral resolution of 60) spanning 3500–9500 \AA~for thousands of galaxies conducted by 2.5m Javalambre Survey Telescope (JST250) telescope at the Astronomical Observatory of Javalambre (OAJ), Teruel, Spain. Its wavelength coverage can detect Ly$\alpha$ emission across a wide redshift range of approximately z = 2 to 7. Studying Ly$\alpha$ emitters over this large cosmological volume enables to study galaxy evolution from the early stages to cosmic noon across diverse environments, including large-scale overdensities, cosmic filaments, providing insights into their role within the cosmic web. In this letter we present the discovery of an extended bright Ly$\alpha$ nebula (named as "RaJav") associated with a quasar (J1620+4334) at z=2.245 using J-PAS. The J-PAS images used for this study were taken from the J-PAS Early Data Release (EDR), released in November 2024, covering 17 deg$^{2}$ with all the J-PAS filters (Vázquez Ramió et al. in prep.). 
\vspace{-0.5cm}
\section{Results}
\subsection{Identification of quasar JPAS-9600-10844}
From the systematic search (see Appendix \ref{sec:AppA1}) for extended Ly$\alpha$ nebulae around spectroscopically confirmed quasars at z$>$2 in J-PAS EDR, we detected an extended object (label R2 in Fig.\ref{Fig:jpasim}a) near to the bright quasar SDSS J162029.07+433451.1 (label R1; hereafter J1620+4334) with a similar redshift in the  J-PAS photospectrum (J-spectrum; Fig. \ref{Fig:jspec}). The better spatial resolution and depth of HSC broad band images are able to resolve the central compact core (hereafter JPAS-9600-10844). The J-spectra of JPAS-9600-10844 clearly shows the peaks of Ly$\alpha$ $\lambda$1216, CIV $\lambda$1548, and HeII $\lambda$1640 emission lines and other peaks of CIII] $\lambda$1909 and MgII] $\lambda$2803 are consistent with the spectroscopic redshift of the quasar (z=2.245). We estimated a photometric redshift of 2.265 $\pm$ 0.021 using LeMoNNADE (Hernán-Caballero et al., in prep.). The center of these objects are separated by a distance of 7.31\as (60.2 kpc). The presence of broad emission lines in both SDSS spectra and J-spectra indicates that these are type 1 quasars. The LOFAR radio observation of this field shows the radio emission extended from the J1620+4334 to JPAS-9600-10844 at 2$\sigma$ (contour of radio emission is shown in Fig.~\ref{Fig:jpasim}d overlaid on i-band image of HSC). The logarithm of the ratio of radio (LOFAR) flux to the NIR (WISE) flux is 0.63 implies it is a radio quiet quasar \citep[][D. Fernández Gil et al. in prep.]{2023Hardcastle}.
\vspace{-0.5cm}
\subsection{Extended nebular emission}
We investigated the spatial distribution of the extended object in Ly$\alpha$, CIV, HeII, CIII] and MgII] emission lines covering J-PAS images. The three filters J0390, J0400, and J0410 covering Ly$\alpha$ emission at z=2.245 show a clear detection with signal-to-noise ratios (S/N) of 9.65, 9.28, and 5.1, respectively, while the extended CIV emission is detected at 4.2$\sigma$ in J0500 and 7.3$\sigma$ in J0510. The depth of the J0410 image is $\sim$ 3 $\times 10^{-16} $ erg s$^{-1}$ cm$^{-2}$ arcsec$^{-2}$ at 2$\sigma$. We co-added three adjacent filter images covering Ly$\alpha$ and CIV lines to improve the S/N (Ly$\alpha$ SB limit increased to $\sim$ 1.86$ \times 10^{-16} $ erg s$^{-1}$ cm$^{-2}$ arcsec$^{-2}$) as shown in Fig.~\ref{Fig:jpasim}a-b (see Appendix \ref{sec:AppA2}). The bridge connecting two quasars is detected at 2$\sigma$ in Ly$\alpha$ and 4.95 $\sigma$ in CIV, the region R3 shows a 3.8$\sigma$ in Ly$\alpha$ but only 1.5 $\sigma$ in CIV. The total end to end size of the underlying extended emission (named as "RaJav") in 2$\sigma$ isophotal area (contour in Fig.~\ref{Fig:jpasim}a) of the coadded Ly$\alpha$ image is 15.38\as (126.7 kpc) and the total Ly$\alpha$ and CIV luminosity are: $5.8 \pm 0.7 \times 10^{44}$ erg s$^{-1}$  and $3.7 \pm 0.5 \times 10^{44}$ erg s$^{-1}$ respectively after subtracting the continuum and contribution of both quasars (see Appendix \ref{sec:AppA3}). By constructing a 3D data cube for J-PAS enabled IFU-like analysis of the nebula. The bridge connecting the two quasars show emission in the J0390 and J0400 Ly$\alpha$ filters, while CIV is detected across three filters (J0490, J0500, and J0510), indicating a broader line profile (see Fig. \ref{sec:AppA3}). Around J1620+4334, we observe spatial variations in both the strength and width of these lines. Notably, CIV emission becomes stronger toward the north, and at the quasar's edge, Ly$\alpha$ and CIV exhibit comparable strengths. See Appendix~\ref{sec:AppA4} for the SB profile of RaJav.
\vspace{-0.5cm}
\subsection{Continuum emission}
To trace the continuum emission around the JPAS-9600-10844, we analyzed J-PAS filters covering the continuum, along with the broader J-PAS iSDSS bands. The continuum emission is too faint to be detected in single J-PAS NB filters, but the deeper iSDSS allows detection at 12.39$\sigma$. Additionally, we examined the broad-band g, r, i, y, and z images from HSC. An extended emission detected in the blue filters and brighter at shorter wavelengths, with the g-band showing the strongest detection at 20$\sigma$ (Fig.~\ref{Fig:jpasim}b). The continuum emission is extending 3.1 \as (25.4 kpc) in the direction of J1620+4334 from the center of JPAS-9600-10844.
The green contours in Fig.~\ref{Fig:jpasim}b are from HSC g-band overlaid on the J-PAS co-added image of CIV + cont. The g-band filter covers Ly$\alpha$ partially and fully covers CIV and HeII emission lines. J-spectrum also shows a significant blue continuum on the extended region. 
\begin{figure}
\centering
\includegraphics[angle=0,width=9cm]{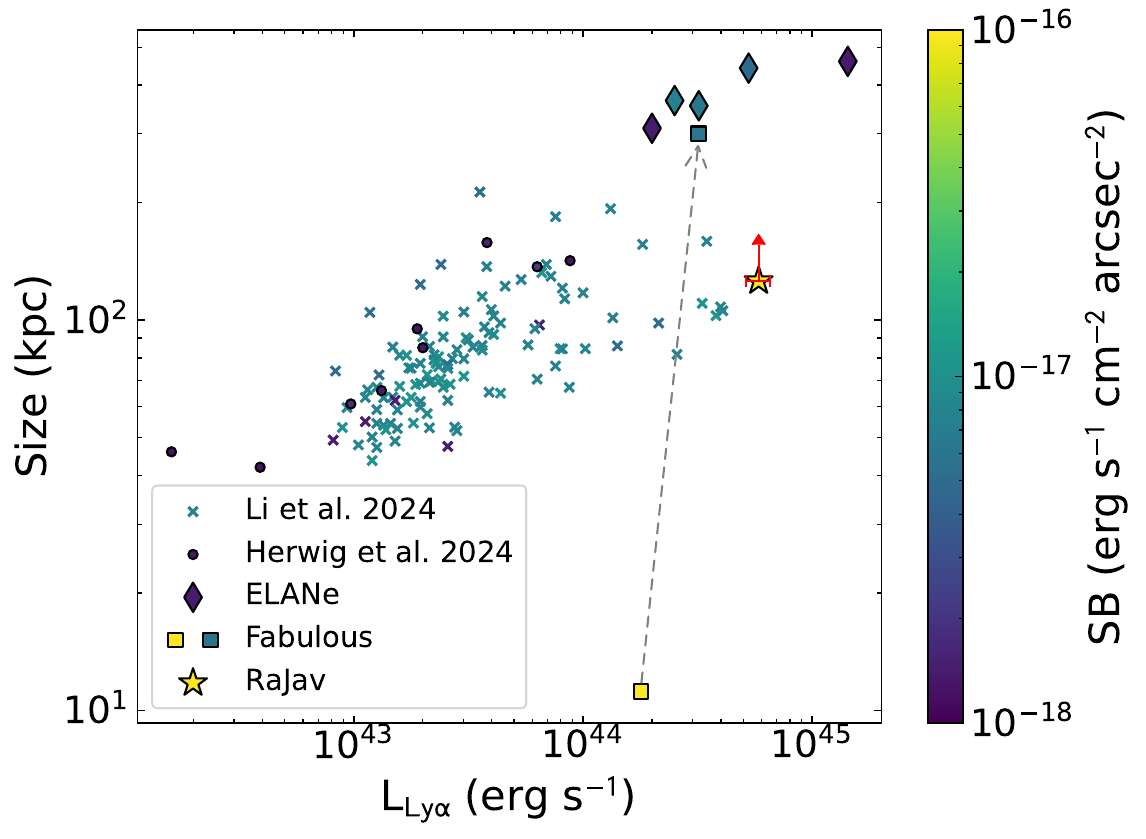}
\caption{Luminosity-size diagram of Ly$\alpha$ nebulae: from \cite{2024Li}, around quasar pairs \citep{2024Herwig}, ELANe \citep{2014Cantalupo,2015Hennawi,2017Cai,2022Arrigoni, 2018Arrigoni}, and RaJav (yellow star). Colors show the SB depth (2$\sigma$) of each nebula. Fabulous nebula is also shown in J-PAS depth (green square).}
\label{Fig:lumsz}
\end{figure}
\vspace{-0.4cm}
\section{Discussion} 
\subsection{Luminosity–size relation}
Previously discovered gigantic ELANe were detected with deep observation of SB $>10^{-18} $ erg s$^{-1}$ cm$^{-2}$ arcsec$^{-2}$, and have Ly$\alpha$ luminosity above 10$^{44}$ erg s$^{-1}$  and extended physical size (> 100 kpcs) mostly associated with multiple quasars in the over-dense environments \citep{2014Cantalupo, 2015Hennawi, 2018Cai, 2023Li, 2024Li}. The discovered diffuse faint extended emission in Ly$\alpha$ and CIV connecting the J1620+4334 and JPAS-9600-10844 quasars is most likely a new ELAN, RaJav, with a physical size (end to end) of $\sim$ 15.38 \as (126.7 kpc) and a Ly$\alpha$ luminosity of $\sim 5.8 \pm 0.7 \times 10^{44}$ erg s$^{-1}$ in the 2$\sigma$ SB contour. The spatial extents and Ly$\alpha$ luminosity of nebulae strongly depend on the limiting SB \citep{2021Kimock}. 
The comparison of luminosity–size relation of RaJav with other Ly$\alpha$ nebulae in the literature is given in Fig.~\ref{Fig:lumsz}. At same SB limit, there is strong relation between the luminosity and size of these Ly$\alpha$ nebulae \citep{2023Arrigoni}. However, our nebula in this sample stands as a clear outlier, being exceptionally large and luminous at shallow J-PAS SB. For direct comparison, we included the luminosity–size relation of Fabulous nebula from \citet{2018Arrigoni} at the SB limit of J-PAS (yellow square) which clearly demonstrates the larger size difference of our nebula at the same SB cut. Comparing this, at the same SB cutoff as Fabulous ELANe, the extrapolated Rajav has size of >1Mpc, potentially the largest ELAN at SB of $>10^{-18} erg s^{-1} cm^{-2} arcsec^{-2}$.
\vspace{-0.7cm}
\subsection{Origin of extended nebular emission}
At a distance of 60 kpc from a quasar, the extended emission at the same redshift likely belongs to the quasar's CGM. If it is a cloud of ionized gas present near a bright quasar, the UV radiation from the quasar photo-ionizes the neutral gas in the cloud, leading to recombination and the emission of strong lines such as Ly$\alpha$, H$\alpha$, C IV, and He II \citep{1991Heckman, 2013Humphrey}. The resolved compact object inside this nebula with the spectral features of an AGN with broad emission lines indicates the presence of a faint quasar (JPAS-9600-10844). 

Additionally, the extended CIV emission (quasar emission subtracted) with luminosity of $3.7 \pm 0.5 \times 10^{44}$ erg s$^{-1}$ indicates that the ionized CGM gas is metal rich \citep{2019Marques-Chaves}. This process might not be homogeneous, given the spatial variation in the strength of CIV emission. Outflows or winds from the quasar may interact with the surrounding gas, producing shocks that excite emission lines through collisional excitation and ionization \citep{2001Taniguchi, 2006Mori} also lead for the extended emission in Ly$\alpha$, CIV and HeII. The combination of double-peaked, strong blue peak asymmetry with broad Ly$\alpha$ and CIV spectral (SDSS) profile of J1620+4334 also strongly suggest that this quasar is undergoing powerful outflows (v$>$3000 km/s). The detected Ly$\alpha$ and CIV extended emission in contiguous multiple filters consistent with photo-ionization by the quasar outflows. The extended radio emission might be the non-thermal synchrotron emission from quasar driven jet interactions, consistent with the shock influence, or may originate separately from two quasars. Although the presence of infalling gas is not discarded, the chemical enrichment and the kinematics suggest that outflows are the dominant mechanisms in this system. The spatial variation in the width and strength of the Ly$\alpha$ and CIV line might be because of the contribution of different powering mechanisms.
 
The presence of a broad Ly$\alpha$ emission line makes gravitational cold accretion a less likely primary powering mechanism, and resonant scattering alone can only contribute for nebular sizes up to about 10 kpc \citep{2021Mitchell}. Based on current observations, we suggest that a combination of outflows and photoionization is the dominant mechanism powering the extended nebula. However, we cannot entirely rule out the possibility that a combination of gravitational cold accretion and resonant scattering also contributes to the observed emission. To confirm this and study the detailed kinematic properties of the ionized gas nebulae we require deep IFU observations. A large scale structure of Ly$\alpha$ emitters (LAE) at redshift range between 2.2 to 2.3 is reported in \cite{2023Zhang}. Our nebula is also located in the large scale structure of quasars over-density of 88 quasars (or LAE) in the comoving volume of (129 cMpc)³.
\vspace{-0.3cm}
\subsection{ Quasar positive feedback or galaxy interaction}
Galaxies at z = 1 to 3, corresponding to most important and active epochs of cosmic Star formation (SF), are crucial for understanding the co-evolution of galaxies and supermassive black holes \citep{2003Dickinson, 2020Florez}. During this period, both SF and AGN activity peaked and AGN feedback is thought to play a pivotal role in regulating SF, either by heating or expelling gas, thus shaping the galaxy’s evolution \citep{2024Zhao}. Ly$\alpha$ nebulae have been observed around star-forming galaxies \citep{2000Steidel, 2019Li}, where photoionization from young stars plays a key role. 

In addition to compact emission, JPAS-9600-10844 shows significant extended emission in HSC g, r, and i bands (brightest in g), with J-spectra revealing rest-frame UV continuum indicative of intense SF. These broad bands also cover the UV emission lines such as Ly$\alpha$ (partially), C IV and He II in g band, and O III] in i-band may contribute to the emission. This suggests SF, potentially triggered by quasar-driven shocks or interaction of host galaxy with J1620+4334. J-PAS IFU-like analysis reveals CIV emission in single filter in this area consistent with ionization from young stars in HII regions. This result suggests that the photoionization from both AGN activity and SF contribute to powering the Ly$\alpha$ nebula, making this massive ELAN a valuable laboratory for exploring their interplay in galaxy evolution.
\vspace{-0.8cm}
\section{Conclusions}
This work presents the discovery of RaJav, a large and massive Ly$\alpha$ nebula, with two embedded quasars at the same spectroscopic redshift of z=2.245 in the wide area NB survey of J-PAS EDR. The end to end size (> 125 kpcs) with a Ly$\alpha$ luminosity of $\sim5.8 \pm 0.7 \times 10^{44}$ erg s$^{-1}$ at $1.86 \times 10^{-16} $ erg s$^{-1}$ cm$^{-2}$ arcsec$^{-2}$ (2$\sigma$ isophote) suggest that this is one of the most extreme systems identified to date at this SB cut. Although the current observational depth limits confirmation of its full extent, it stands out compared to previously reported ELANe, and may be the largest identified at a similar depth. The high-resolution of HSC g-band image and J-spectra reveal a second hidden quasar located 60.2 kpcs from the bright J1620+4334.
The nebula exhibits broad Ly$\alpha$ emission line and signs of metal enrichment, indicative of outflow-driven shocks and photoionization as key ionization mechanisms. The filamentary structures connecting two quasars and the location in the over dense region suggests a physical link to the large-scale cosmic web.
This discovery sheds lights on CGM gas flows and accretion process during the peak epoch of galaxy formation. The coexistence of AGN activity and star formation in this complex environment supports a scenario in which quasar-driven feedback plays a crucial role in regulating galaxy evolution. Future observations are planned for a more detailed follow-up integral field spectroscopic study of this object.
\begin{acknowledgements}
RPT thanks the European Union - NextGenerationEU through the
Recovery and Resilience Facility program (RRF) Planes Complementarios
con las CCAA de Astrof\'{\i}sica y F\'{\i}sica de Altas Energ\'{\i}as -
LA4. This paper has gone through internal review by the J-PAS collaboration. RPT thanks Ana and Ivan for their suggestions. Based on observations made with the JST250 telescope and JPCam camera for the J-PAS project at the OAJ, in Teruel, owned, managed, and operated by CEFCA. We acknowledge the OAJ Data Proc. \& Arch. Unit (UPAD) for reducing and calibrating the OAJ data used in this work. Funding for the J-PAS Project has been provided by the Governments of Spain and Arag\'on through the Fondo de Inversiones de Teruel; the Aragonese Government through the Research Groups E96, E103, E16\_17R, E16\_20R, and E16\_23R; the Spanish Ministry of Science and Innovation (MCIN/AEI/10.13039/501100011033 y FEDER, Una manera de hacer Europa) with grants PID2021-124918NB-C41, PID2021-124918NB-C42, PID2021-124918NA-C43, and PID2021-124918NB-C44; the Spanish Ministry of Science, Innovation and Universities (MCIU/AEI/FEDER, EU) with grants PGC2018-097585-B-C21 and PGC2018-097585-B-C22; the Spanish Ministry of Economy and Competitiveness (MINECO) under AYA2015-66211-C2-1-P,
AYA2015-66211-C2-2, and AYA2012-30789; and European FEDER funding
(FCDD10-4E-867, FCDD13-4E-2685); the Brazilian agencies FINEP, FAPESP,
FAPERJ and by the National Observatory of Brazil. Additional funding was
provided by the Tartu Obs. and by the J-PAS Chinese Astron. Consortium. We acknowledges support from MICIU/AEI/10.13039/501100011033 for the projects: PID2023-149420NB-I00 (SDG), PID2023-147386NB-I00 (RA and AH), PID2022-141755NB-I00 (LADG, YJT, RGD), PID2022-136598NB-C32 (YJT) and Severo Ochoa grant CEX2021-001131-S (LADG, YJT, RGD). SDG thanks PROMETEO CIPROM/2023/21 (Generalitat Valenciana) and Project (VAL-JPAS) reference ASFAE/2022/025 (PRTR 2022, MICIU, EU NextGenerationEU and GV).
\end{acknowledgements}
\vspace{-1.2cm}
\bibliography{ref}{}
\bibliographystyle{aa}

\twocolumn
\begin{appendix} 
\section{Further analysis of the nebula}
\label{sec:AppA}
\subsection{Pipeline and object selection}
\label{sec:AppA1}
We have developed a dedicated pipeline (Rahna et al., in prep.) for the systematic identification of extended Ly$\alpha$ nebulae around quasars and galaxies at redshifts z $>$ 2 using J-PAS observations.
The J-PAS survey is capable of detecting the Ly$\alpha$ emission line over the redshift range z=2 to z=7. The spectroscopic redshift of the objects are taken from DESI, SDSS, and J-PAS photospectra (LeMoNNADE). Learning Model with Neural Network Adaptive Distance Estimator (LeMoNNADE) is a tool to estimates photometric redshifts and performs spectral classification of J-PAS sources based on their SEDs (Hernan-Caballero et al. in prep.). The pipeline select the nebula candidates based on the presence of spatially extended Ly$\alpha$ emission around quasars and galaxies that exceeds the point spread function (PSF) in the corresponding J-PAS narrow-band filter. We also analyze the nebula in other prominent emission line images (e.g., CIV, HeII, CIII, and MgII) to understand their physical origin. The physical extent of each nebula is quantified by measuring its size at the 2$\sigma$ SB isophote. The pixel scale of J-PAS is 0.419"/pixel.Throughout this paper, we adopt an angular scale of 8.238 kpc/" at the redshift of the source (z = 2.245), assuming $\Lambda$CDM cosmology with $H_0 = 70 \mathrm{km \, s^{-1} \, Mpc^{-1}} , \Omega_m = 0.3$, and$~\Omega_\Lambda = 0.7$.

\begin{figure*}[ht]
\centering
\includegraphics[angle=0,width=18.5cm]{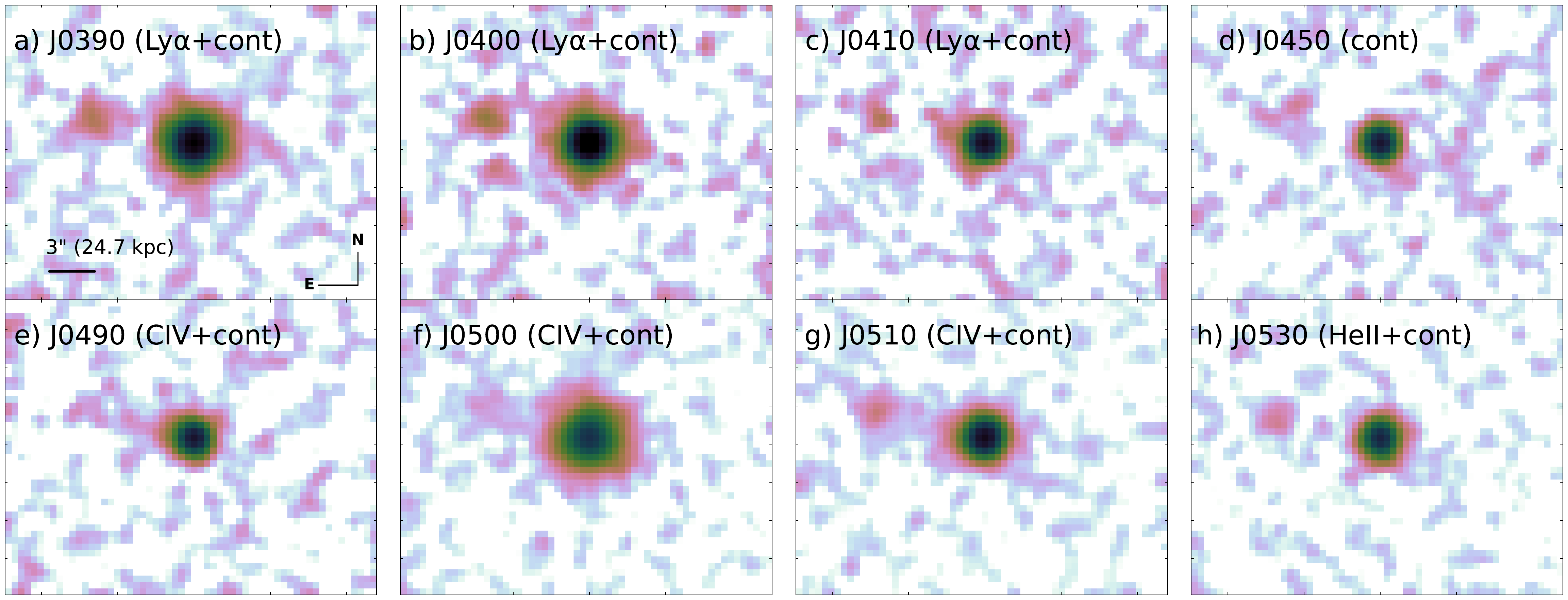}
\caption{J-PAS NB image of a) J0390, b) J0400, and c) J0410 covering Ly$\alpha$+cont, J0450 covering continuum e) J0490, f) J0500, g) J0510, covering CIV+cont, and h) J0530 covering HeII+cont.}
\label{Fig:jpasimgs}
\end{figure*}

In our search for extended Ly$\alpha$ nebulae within the J-PAS EDR sample, we identified the object near the quasar J1620+4334 that is the focus of this study. The spectroscopic redshift (z=2.245) of the J1620+4334 is taken from DESI spectra. At this redshift the Ly$\alpha$ of this quasar covers three J-PAS narrow-band filters: J0390 (centered at 3904 $\AA$), J0400 (centered at 3996 $\AA$), and J0410 (centered at 4110 $\AA$). Fig. \ref{Fig:jpasimgs} shows the detected extended emission in Ly$\alpha$, CIV, and HeII covering J-PAS NB images. We also detected extended object in HeII at 6.52$\sigma$ in J0530, and CIII] at 3$\sigma$ In J0620.
The depth of the J0410 filter is (23.5 $mag~ arcsec^{-2}$) or 3.7$ \times 10^{-16} $ erg s$^{-1}$ cm$^{-2}$ arcsec$^{-2}$ and the broad band filter iSDSS is (24.6 $mag~arcsec^{-2}$) or 3.3 $ \times 10^{-16} $ erg s$^{-1}$ cm$^{-2}$ arcsec$^{-2}$ (5 $\sigma$ in 1 $arcsec^{2}$ aperture). 

We also looked to wide Hyper Suprime-Cam DR2 \citep[HSC;][]{2019Aihara} survey optical imaging in g, r, i, z, and y broad band filters offers high spatial resolution images (0.79\as in g band) with better sensitivity. Gnuastro \citep{2015Akhlaghi, 2019Akhlaghi} routines \texttt{NoiseChisel}, \texttt{Segment}, and \texttt{MakeCatalog} \citep{2019Akhlaghi} are used to estimate the S/N by segmentation. The end-to-end size of the nebula in the 2$\sigma$ SB contour of g band (green contour in Fig. \ref{Fig:jpasim}b) is 16.49\as (135.84 kpcs). A radio counterpart of this object was detected by LOFAR \citep[Low-Frequency Array;][]{2013Haarlem} Two-Metre Sky Survey at 177 MHz (red contour in Fig. \ref{Fig:jpasim}d). The spatial resolution of LOFAR is 6\as. 

\subsection{Image stacking}
\label{sec:AppA2}
We stacked three filters of J-PAS Ly$\alpha$ (J0390, J0400,J0410) to get high S/N. The images are converted to SB units (erg s$^{-1}$cm$^{-2}$ arcsec$^{-2}$) for the co-addition.
The SB limit of the images are calculated using Gnuastro routines. We convolved each image to the worst PSF (1.39" for Ly$\alpha$ and 1.25" for CIV) among the three filters before stacking. The CIV stacked image was created using J0490 (centered at 4902 $\AA$), J0500 (centered at 5002 $\AA$), J0510 (centered at 5097 $\AA$), while the continuum image was created using J0420 (centered at 4203 $\AA$), J0440 (centered at 4403 $\AA$), and J0450 (centered at 4503 $\AA$) filters. We also noticed there is an excess of HeII emission in the J-Spectrum of JPAS-9600-10844 as compared to the bright quasar spectrum. However high resolution spectra is necessary to give a conclusion. The identified nebula, RaJav includes R1, R2, and R3 regions as in Fig. \ref{Fig:jpasim}a within the 2$\sigma$ Ly$\alpha$ SB contour. We have nicknamed the nebula "RaJav" meaning "King" in Malayalam.
\subsection{Luminosity calculation}
\label{sec:AppA3}
The total Ly$\alpha$ and C IV luminosities for the J1620+4334 and JPAS-9600-10844 are calculated by summing the continuum-subtracted emission from three adjacent filters. The average flux of the three filters around Ly$\alpha$ and CIV far from the emission lines are used to the estimation of the continuum.
The aperture for the flux calculation in the nebula was defined by 2 sigma isophotal contour in the Ly$\alpha$ image, and continuum subtraction was performed by measuring the flux within this contour in the continuum image and subtracting it from the line+continuum flux in the Ly$\alpha$ image. To remove contamination from the central quasars, we estimated their contribution using a circular aperture defined by the point spread function (PSF) in the Ly$\alpha$ filter image, which has a full width at half maximum (FWHM) of 1.39". A radius equal to the FWHM was adopted. We also applied the aperture corrections through a curve-of-growth analysis of bright stars in the field for each filter used in the luminosity measurements to ensure that the extended wings of the PSF were properly accounted for. The Ly$\alpha$ fluxes for two quasars were subtracted from the total Ly$\alpha$ flux of the nebula within the 2$\sigma$ isophote. 
The fluxes are corrected for the Milky Way extinction and also for the overlapping of 3 adjacent filters of J-PAS.
The center of the new quasar is taken from the resolved image of HSC- g. 
The total Ly$\alpha$ luminosity of J1620+4334 is $2.14 \pm 0.6 \times 10^{45}$ erg s$^{-1}$ and $4.37 \pm 0.5 \times 10^{44}$ erg s$^{-1}$ for CIV . For the JPAS-9600-10844, the total luminosities are $1.06 \pm 0.1 \times 10^{44}$ erg s$^{-1}$ for Ly$\alpha$ and $3.31 \pm 0.8 \times 10^{43}$ erg s$^{-1}$ for CIV.
After subtracting the continuum and contribution of both quasars, the total Ly$\alpha$ and CIV luminosity of RaJav nebula are $5.84 \pm 0.7 \times 10^{44}$ erg s$^{-1}$ and $3.7 \pm 0.5 \times 10^{44}$ erg s$^{-1}$ respectively.

\subsection{Surface brightness profile}
\label{sec:AppA4}

\begin{figure*}
\centering
\begin{minipage}{0.49\textwidth}
\centering
\begin{tikzpicture}
\node at (0cm,0cm)
       {\includegraphics[width=0.96\linewidth]{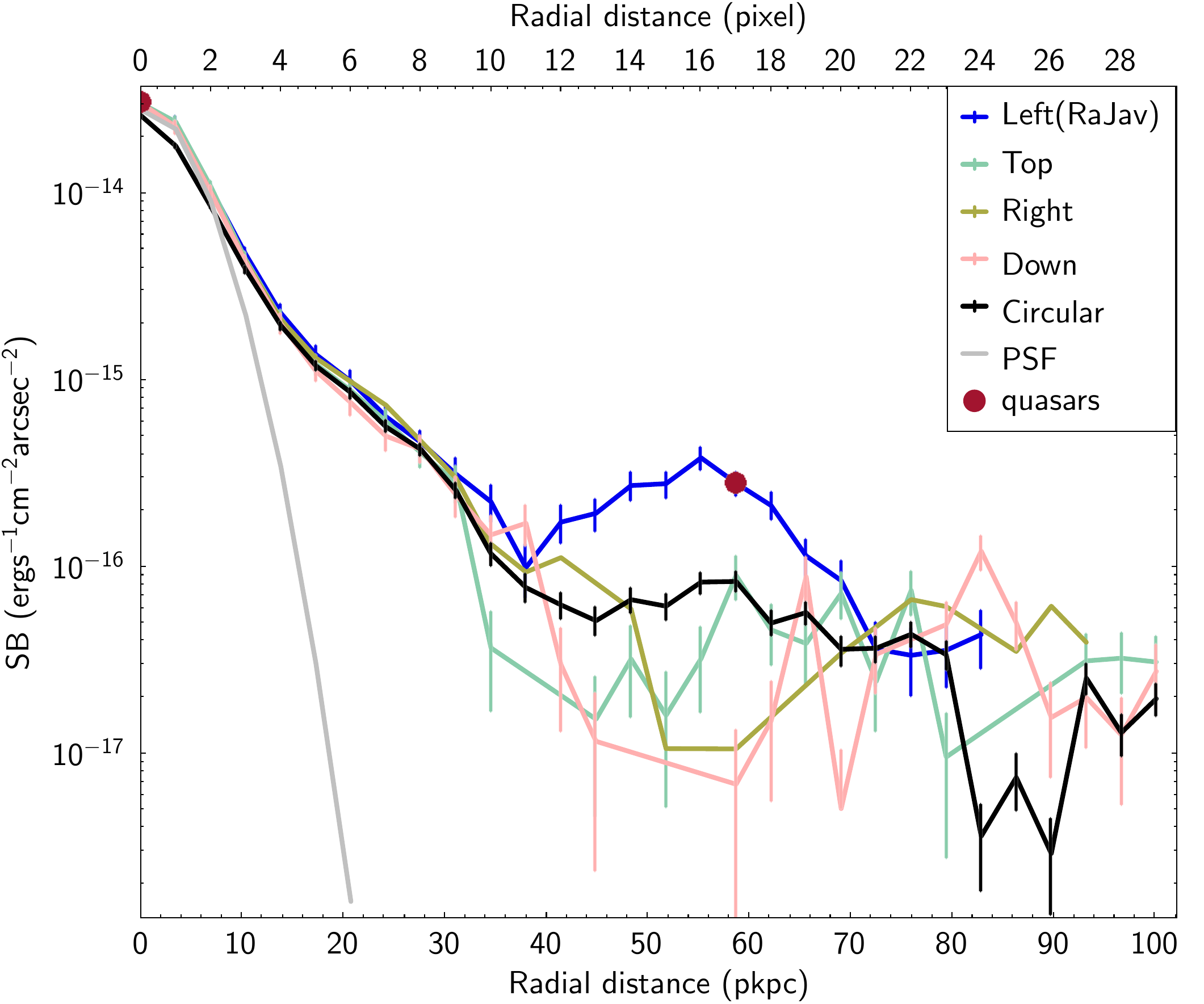}};
\node at (1.5cm,2cm)
       {\includegraphics[width=0.15\linewidth]{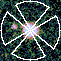}};
\end{tikzpicture}
\end{minipage}
\hfill
\begin{minipage}{0.49\textwidth}
\centering
\begin{tikzpicture}
\node at (0cm,0cm)
       {\includegraphics[width=0.98\linewidth]{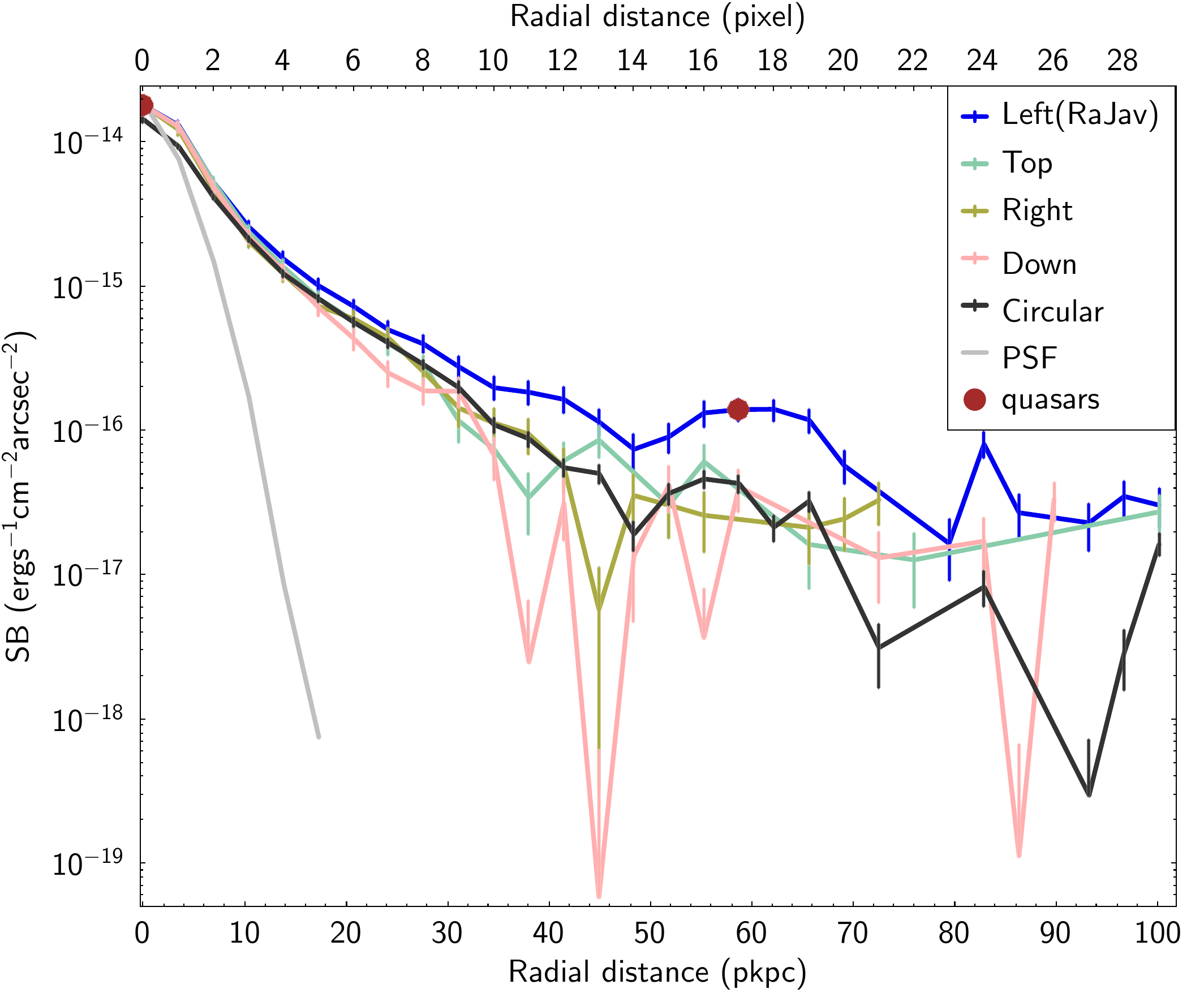}};
\node at (1.5cm,2cm)
       {\includegraphics[width=0.15\linewidth]{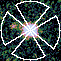}};
\end{tikzpicture}
\end{minipage}

\caption{Radial surface brightness profiles extracted in four angular sectors around J1620+4334 (Left: from coadded Ly$\alpha$+cont, Right: from coadded CIV+cont image). The conical regions (over-plotted on the coadded image is on top right) correspond to position angle intervals of 50$^{\circ}$–120$^{\circ}$ (mint green), 140$^{\circ}$–210$^{\circ}$ (blue), 230$^{\circ}$–300$^{\circ}$ (pink), and 320$^{\circ}$–30$^{\circ}$ (olive). The circularly averaged SB profile are in black. The distances of the quasars from the center of J1620+4334 are indicated by red circles and the grey curve is from the PSF.}
\label{Fig:radprof}
\end{figure*}

To investigate possible azimuthal variations in the environment of J1620+4334, we extracted radial surface brightness profiles in four distinct angular sectors around the quasar in the coadded image of Ly$\alpha$+cont and CIV+cont in Fig. \ref{Fig:jpasim}. Each profile was measured within a conical wedge defined by fixed position angle ranges (measured counter-clockwise from north): 50$^{\circ}$ – 120$^{\circ}$ (mint green), 140$^{\circ}$ – 210$^{\circ}$ (blue), 230$^{\circ}$ – 300$^{\circ}$ (pink), and 320$^{\circ}$ –30$^{\circ}$ (olive) Fig. \ref{Fig:radprof}.
The distance of the quasars from the center of J1620+4334 are labeled in red circle.
In each sector, the flux was azimuthally averaged in concentric annuli to construct the radial profile using Gnuastro routine \texttt{astscript-radial-profile} with \texttt{-{}-azimuth} function \citep{2024Infante}. This approach is particularly suited to revealing directional asymmetries, in particular, it allows us to check for excess extended emission consistent with a nebular component located preferentially to the east (left) of the quasar position. Fig. \ref{Fig:radprof} shows prominent extended Ly$\alpha$ emission at radial distances of 25 - 75 kpcs, after PSF at SB level of 1 $\times$ $10^{-16}$ to 4 $\times$ $10^{-16} $ erg s$^{-1}$ cm$^{-2}$ arcsec$^{-2}$~, when compared with the radial profiles along different directions where the quasar PSF dominates. 
We also included the circularly averaged SB profile (black) as usually done in the literature. Asymmetric profile of the nebula washed out in the circularly averaged profile compared to the wedge profile. Compared to the Ly$\alpha$ SB profiles of other nebulae reported in the literature e.g., the average profile from \citet{2016Borisova} and quasar-pair nebulae \citep{2024Herwig}, this profile (at 25-75 kpcs) is brighter by a factor of $\sim$5–10 and shows a flatter shape. The direct comparison and the detailed analysis on the SB profile would require future deep follow-up IFU observations.

\begin{figure*}[ht]
\centering

\includegraphics[width=0.46\linewidth]{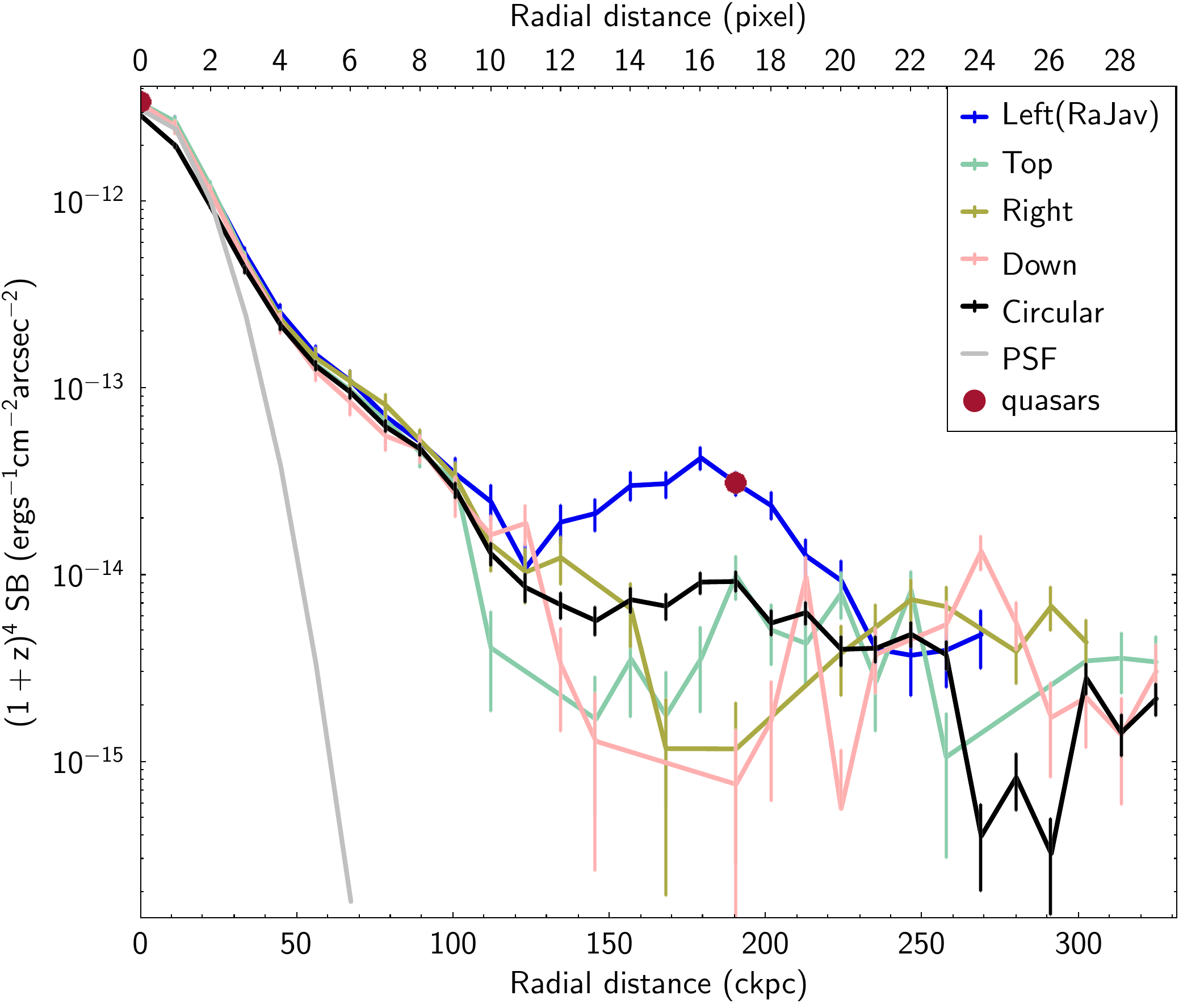}
\includegraphics[width=0.46\linewidth]{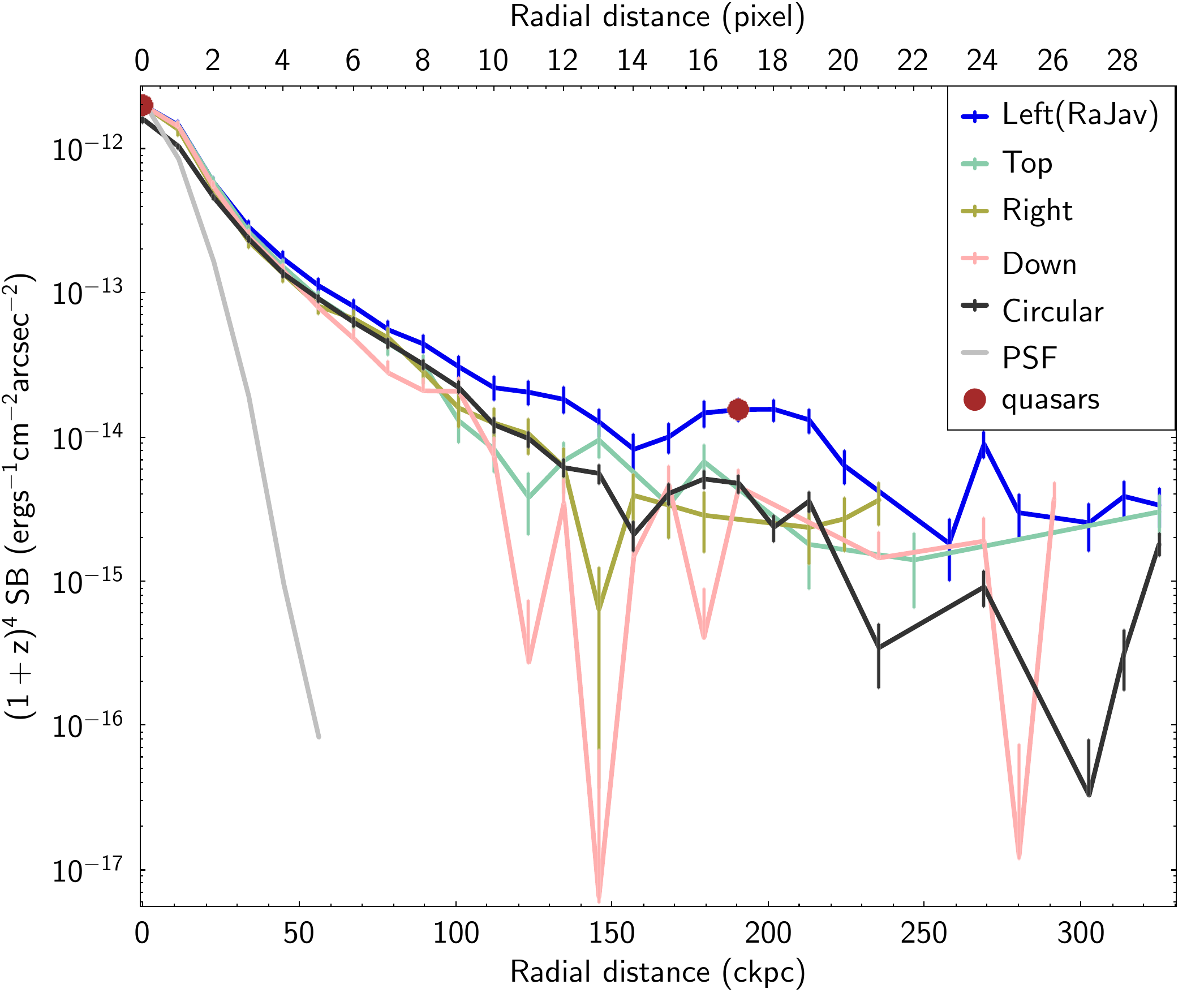}
\caption{Cosmologically corrected radial surface brightness profiles extracted in four angular sectors around J1620+4334 (Left: from coadded Ly$\alpha$+cont, Right: from coadded CIV+cont image). The conical regions correspond to position angle intervals of 50$^{\circ}$–120$^{\circ}$ (mint green), 140$^{\circ}$–210$^{\circ}$ (blue), 230$^{\circ}$–300$^{\circ}$ (pink), and 320$^{\circ}$–30$^{\circ}$ (olive). The circularly averaged SB profile are in black. The distances of the quasars from the center of J1620+4334 are indicated by red circles and the grey curve is from the PSF.}
\label{Fig:radprof-cos}
\end{figure*}

A similar result is obtained in CIV profile and the contribution of the bridge is very significant compared to other axis wedge profiles. This extended emission does not belong to the quasar wing profile, but instead shows the rise in the SB above 24 kpcs. The cosmologically corrected SB profiles are given in Fig. \ref{Fig:radprof-cos}.

\end{appendix}
\end{document}